\documentclass[prl,twocolumn]{revtex4-1}

\usepackage[applemac]{inputenc}
\usepackage{amssymb,amsmath}
\usepackage{graphicx}
\usepackage{float}
\usepackage{color}

\newcommand{\ket}[1]{|{#1}\rangle}

\newcommand{\beginsupplement}{%
        \setcounter{table}{0}
        \renewcommand{\thetable}{S\arabic{table}}%
        \setcounter{figure}{0}
        \renewcommand{\thefigure}{S\arabic{figure}}%
     }

\begin{document}	

\title{Microwave Emission from Hybridized States in a Semiconductor Charge Qubit}

\date{24 July 2015}

\author{A.~Stockklauser}
\email{anna.stockklauser@phys.ethz.ch}
\author{V.~F.~Maisi}
\author{J.~Basset}
\altaffiliation{now at Laboratoire de Physique des Solides, Univ. Paris-Sud, CNRS, UMR 8502, F-91405 Orsay Cedex, France,
julien.basset@u-psud.fr}
\author{K.~Cujia}
\author{C.~Reichl}
\author{W.~Wegscheider}
\author{T.~Ihn}
\author{A.~Wallraff}
\author{K.~Ensslin}
\affiliation{Department of Physics, ETH Z\"urich, CH-8093 Zurich, Switzerland}

\begin{abstract}
We explore the microwave radiation emitted from a biased double quantum dot due to the inelastic tunneling of single charges. Radiation is detected over a broad range of detuning configurations between the dot energy
levels with pronounced maxima occurring in resonance with a capacitively coupled transmission line resonator. The power emitted for forward and reverse resonant detuning is found to be in good agreement with a rate equation model, which considers the hybridization of the individual dot charge states. 
\end{abstract}

\maketitle

The electronic properties of semiconductor nanostructures have been widely studied using transport measurements \cite{Beenakker1991}. More recently, the use of radio and microwave frequency measurement techniques has enabled and stimulated a new generation of experiments \cite{Wallraff2004, Schoelkopf2008, Girvin2009b} mostly with superconducting qubits.
In similar device geometries the coupling of double quantum dots in carbon nanotubes \cite{Delbecq2011}, gate-defined GaAs heterostructures \cite{Frey2012,Toida2013}, InAs nano-wires \cite{Petersson2012a}, and graphene \cite{Deng2013a} to GHz frequency coplanar waveguide resonators has been explored. 
Earlier experiments have investigated photon emission and lasing effects in superconducting circuits \cite{Astafiev2007,Hofheinz2011}. More recently experiments studied photon emission from biased double quantum dots. Gain and micromaser action have been predicted \cite{Jin2011} and observed by pumping a single microwave resonator mode \cite{Liu2014f, Liu2015}.

Here, we experimentally explore inelastic tunneling in a semiconductor double quantum dot capacitively coupled to a transmission line resonator by detecting the microwave radiation emitted in the process. The detection of the weak microwave signals is facilitated by the use of near quantum limited parametric amplifiers \cite{Yurke2006, Castellanos2007}. Previous experiments on quantum dots coupled to microwave resonators \cite{Delbecq2011,Frey2012,Toida2013,Petersson2012a,Deng2013a} detected mostly polarizability allowing to extract charge stability diagrams. Here, we demonstrate that the level separation of hybridized states in a double quantum dot, if on resonance with the microwave resonator, can be investigated with high precision by directly detecting inelastic transitions. In particular, the tunneling rates, which are typically difficult to measure in pure dc transport experiments, are directly reflected in the power of the respective emitted microwave radiation. Future options include the possibility to characterize the classical and quantum properties of radiation emission from semiconductor nanostructures at microwave frequencies using correlation function measurements \cite{Bozyigit2011} or state tomography \cite{Eichler2012}.

\begin{figure}
	\centering
\includegraphics[width=8.6 cm]{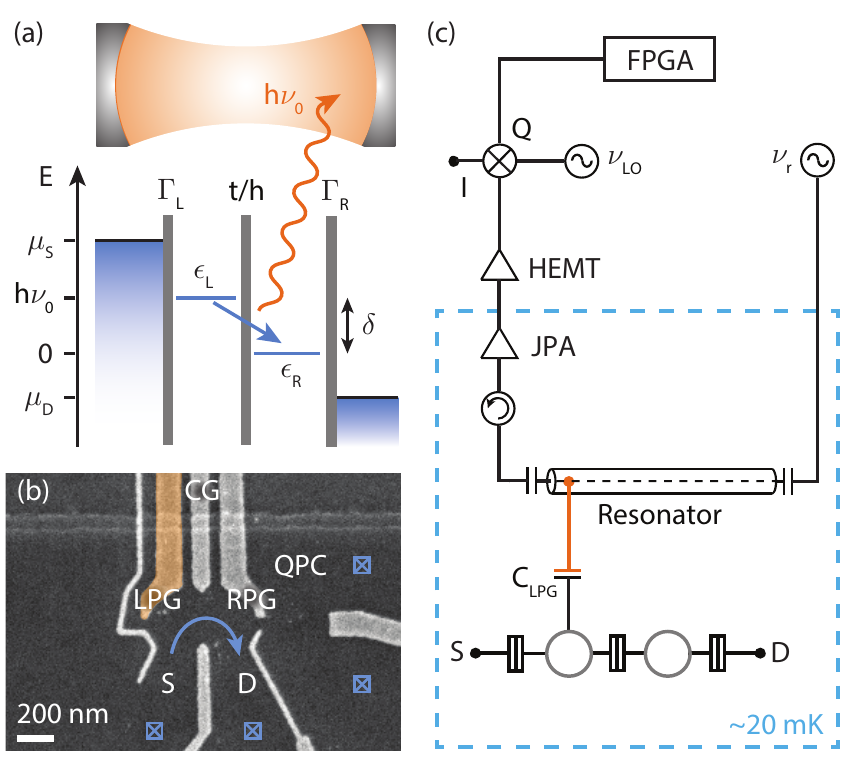}
	\caption{
(a) Energy level structure of a double quantum dot with levels $\epsilon_L$ and $\epsilon_R$. The dots are tunnel coupled to source (S) and drain (D) with chemical potentials $\mu_S$ and $\mu_D$ at rate $\Gamma_\text L$ and $\Gamma_\text R$. The interdot coupling rate is given by $t/h$. We consider the creation of photons in a cavity of resonance frequency $\nu_0$ in dependence on 
the interdot detuning $\delta = \epsilon_L - \epsilon_R$. 
In the depicted situation ($\delta>0$) the detuning is in forward direction with respect to the source-drain bias.
(b) Scanning electron micrograph of the quantum dot structure: center gate (CG), left and right plunger gate (LPG, RPG), quantum point contact (QPC).
The LPG shaded in orange capacitively couples the left dot to the resonator ($C_\text{LPG}$).
(c) Schematic of sample and measurement setup. A coherent microwave signal ($\nu_r$) is applied to the resonator.
The output signal passes a circulator and is amplified by a Josephson parametric amplifier (JPA) and a high electron mobility transistor (HEMT) amplifier. It is mixed with a local oscillator signal ($\nu_\text{LO}$) in a heterodyne detection scheme and processed with a field programmable gate array (FPGA).}
\label{fig:sample}
\end{figure}

The hybrid device explored in our experiments is realized by capacitively coupling a gate-defined double quantum dot to an on-chip superconducting coplanar waveguide resonator [Fig.~\ref{fig:sample}]. 
The gate structure depicted in Fig.~\ref{fig:sample}(b) is patterned on top of a GaAs/AlGaAs heterostructure with a two-dimensional electron gas (2DEG) 90~nm below the surface. The quantum dots are formed by negatively biasing the top gates. The left and right plunger gates (LPG, RPG) control the electrochemical potentials ($\epsilon_\text L, \epsilon_\text R$) of the electrons in the respective dots and thereby the detuning energy $\delta=\epsilon_\text L-\epsilon_\text R$ between the dot levels. 
The center gate is used to adjust the interdot tunnel coupling energy $t$. The tunnel coupling leads to a hybridization of the left and right dot states. The resulting bonding and antibonding states form a two-level system with an energy separation $h\nu_\text q=\sqrt{(2t)^2+\delta^2}$, where $h$ is Planck's constant and $\nu_\text q$ denotes the transition frequency of the charge qubit \cite{Hayashi2003}.

The 200\,nm thick Al coplanar waveguide resonator has a bare resonance frequency of $\nu_0=6.852$~GHz ($28~\mu\text{eV}/h$) 
and a loaded quality factor of 2058. The center conductor of the resonator extends to the left dot forming the plunger gate shaded in orange, which mediates dipole coupling between double quantum dot and resonator [Fig.~\ref{fig:sample}(c)] \cite{Frey2012,Toida2013}. The sample is similar to the devices presented in Refs.~\onlinecite{Frey2012, Basset2013}.

\begin{figure}
	\centering
\includegraphics[width=8.6 cm]{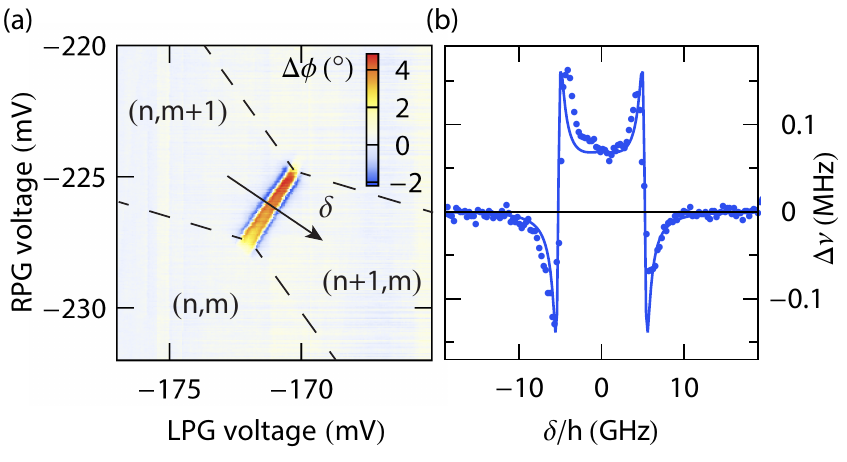}
	\caption{(a) Change in phase $\Delta\phi$ of a constant RF measurement tone applied to the resonator as a function of left and right plunger gate voltages  in the vicinity of the triple points. 
The borders between stable charge configurations with $n$ electrons in the left and $m$ electrons in the right dot are indicated by dashed lines.  The arrow indicates the axis along which the detuning $\delta$ between the dots is varied. A fluctuating background constant in LPG voltage is subtracted.
(b) Associated shift in the resonance frequency $\Delta\nu=\nu-\nu_0$ against detuning $\delta$. The solid line is a Master equation simulation of a Jaynes-Cummings model from which the coupling strength $g/2\pi\approx 13$~MHz, decoherence rate $\gamma/2\pi=250$~MHz and tunnel coupling $2t/h\approx4.4$~GHz are extracted \cite{Frey2012, Toida2013}.
}
	\label{fig:RF}
\end{figure}

We first characterize the properties of the coupled system in microwave transmission measurements as described in Ref.~\onlinecite{Frey2012}. This yields all parameter values relevant for the analysis of the photon emission data presented subsequently. 
A microwave tone of constant frequency $\nu_r$ set to the bare cavity frequency $\nu_0$ is transmitted through the resonator. 
The output field is first amplified by a Josephson parametric amplifier \cite{Eichler2014a} providing quantum limited amplification \cite{Yurke2006, Castellanos2007}.
A high electron mobility transistor amplifier provides a second amplification stage before the signal is mixed with a local oscillator microwave tone.
The signal is subsequently digitally down-converted and processed with a field-programmable gate array, giving access to its  amplitude $A$ and phase $\phi$ [Fig.~\ref{fig:sample}(c)] \cite{Wallraff2004}.

Charge delocalization between the dots leads to a dispersive frequency shift of the cavity seen as a change in phase $\Delta\phi$ of the transmitted tone as plotted in Fig.~\ref{fig:RF}(a). The charge stability regions are indicated together with the $\delta$-axis along which the detuning between the dots is varied \cite{Wiel2002}. In the presented experiments, the number of electrons in each dot is on the order of 10 as determined by quantum point contact charge detection \cite{Field1993, Buks1998}.

The dispersive frequency shift as a function of $\delta$ is measured by recording full transmission spectra.
At detuning energies in the vicinity of the qubit-cavity resonance condition the resonator frequency is strongly shifted [Fig.~\ref{fig:RF}(b)]. 
We calculate the frequency shift numerically using a Jaynes-Cummings model \cite{Childress2004} to extract the resonator-dot coupling strength $g$, the interdot tunnel coupling energy $t$ and the decoherence rate $\gamma$. To analyze the emission experiments performed at different tunnel coupling energies $t$ discussed in the following, we have used a constant value of $g/2\pi \approx 11$~MHz, see supplemental material \cite{SuppMat}. 

With this device, decoherence rates as low as $250\pm50$~MHz were observed, which is more than an order of magnitude lower than for previous samples investigated in our group \cite{Frey2012, Basset2013} and comparable to the lowest rates observed in other experiments \cite{Toida2013, Wallraff2013, Viennot2014a}.
We attribute the improvement to a combination of 
optimized filtering and a very stable 2DEG. It also manifests itself in a reduced electron temperature of $\sim60$\,mK as extracted from 
conductance resonances.
Note that the improved coherence facilitates the emission experiments presented below.

\begin{figure*}
	\centering
\includegraphics[width=17.9 cm]{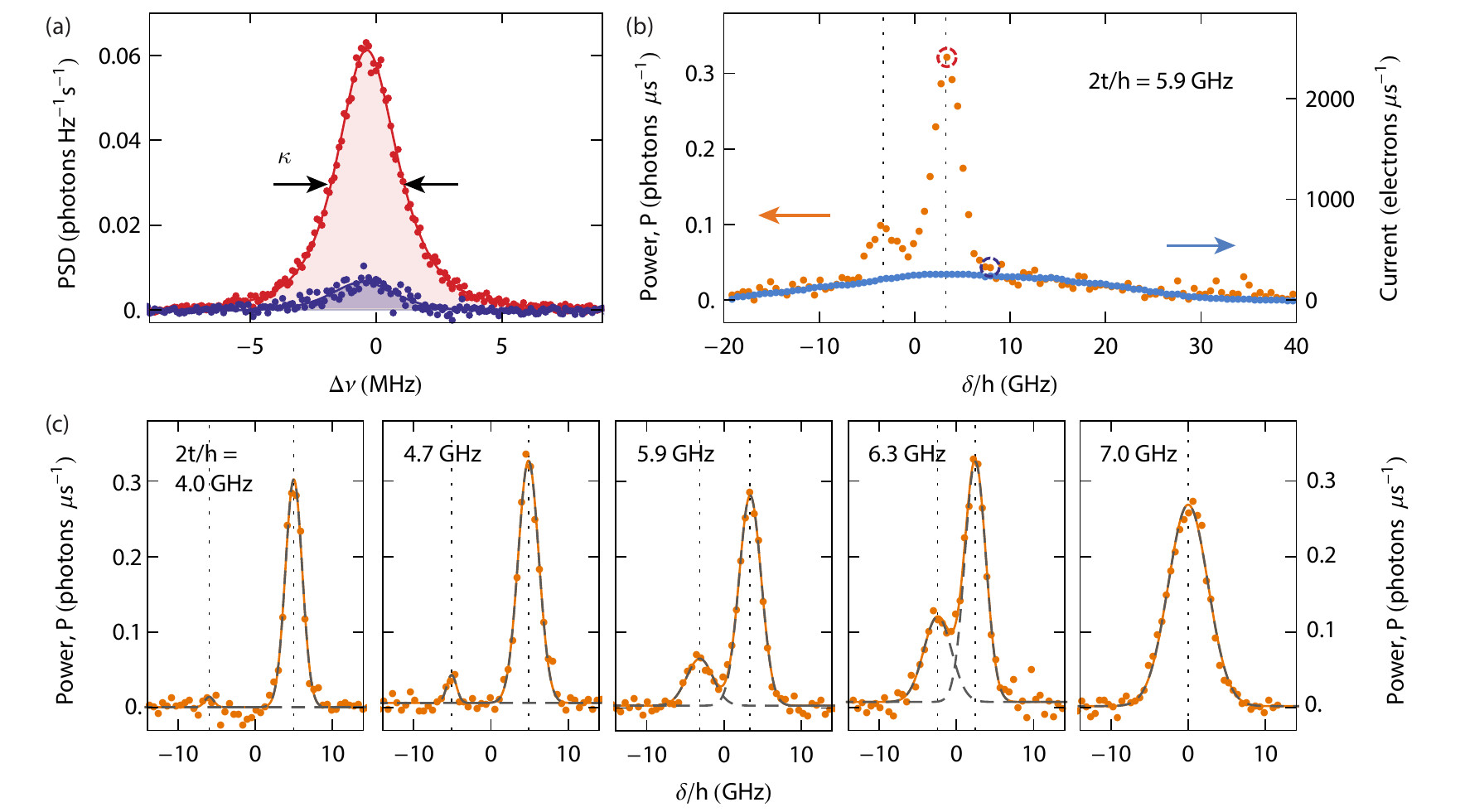} 
	\caption{(a) Power spectral density (PSD) of microwave emission measured at the two detuning energies $\delta$ indicated by dashed circles in (b). The extracted linewidth corresponds to the resonator linewidth $\kappa$.
(b) Plot of the source drain current (blue, scale on the right hand side) and the emission power in units of photons emitted into the cavity per microsecond (orange) against detuning $\delta$. Each point in emission power is obtained by integrating the respective PSD. 
(c) Photon emission power measured for the indicated interdot tunnel rates $2t/h$. The background proportional to the current is subtracted and the emission resonances are fitted using a sum of two Gaussian lineshapes to extract the values of resonant detuning and power (sum: solid orange line; individual lineshapes: dashed gray lines).}
	\label{fig:emission}
\end{figure*}

To study photon emission from the double dot, we apply a bias of $V_\text{SD}=-200~\mu$V ($48~\text{GHz}$) 
between source and drain. The bias is chosen to be smaller than both the charging energies and the single-particle level spacing to ensure single-level transport. The charging energy of each dot is roughly 1~meV ($\approx 200~\text{GHz} \cdot h$), as extracted from the stability diagram, and at $V_\text{SD}=-200~\mu$V transport measurements show no excited states in the bias window. The tunnel coupling to the leads, adjusted to $\Gamma_\text L \approx \Gamma_\text R \sim 1$~GHz, controls the current through the device.

We measure the power spectral density (PSD) of photons emitted from the cavity \cite{Lang2011,SuppMat} with the double quantum dot being the only source of radiation. For detuning values $|\delta/h|$ larger than the resonance frequency, we find a low photon signal of less than $0.01~ \rm{Hz}^{-1}\rm{s}^{-1}$, see purple curve in Fig.~\ref{fig:emission}(a). Close to resonance, the photon signal is significantly increased to more than $0.06~ \rm{Hz}^{-1}\rm{s}^{-1}$ (red curve) corresponding to an average number of $0.015$ photons in the resonator extracted from the integrated PSD. Both data sets fit well to a Lorentzian lineshape (solid lines) taking into account the frequency dependence of the parametric amplifier gain. The linewidths extracted from all PSD measurements are identical to the resonator linewidth $\kappa/2\pi=3.3$~MHz. 

We measure the PSD vs.\ detuning $\delta$ around the lower triple point shown in Fig.~\ref{fig:RF}(a). Integrating Lorentzian fits to the data yields the number of photons emitted from the cavity per unit of time, $P$. The power $P$ is plotted in Fig.~\ref{fig:emission}(b) as a function of detuning $\delta$ (orange) and compared to the simultaneously measured source-drain current (blue). We observe a pronounced resonance in the measured power $P$ at positive detuning ($+\delta_\text{0}$) as well as a lower peak at negative detuning ($-\delta_\text{0}$).  These maxima occur when the energy released in the interdot transition corresponds to the resonator frequency ($\nu_q=\nu_0$) yielding the resonant detuning energies
\begin{align}
\pm\delta_\text{0}=\pm\sqrt{(h\nu_0)^2-(2t)^2} \, .
\label{deltares}
\end{align}

The background emission power away from the resonances is proportional to the current. We associate it with combined photon/phonon processes and tunneling processes between the left or right dot and the continuum of states in the leads. 
The proportionality factor is the same for all measurements and yields a background rate of roughly $1.3\times10^{-4}$ photons emitted from the cavity per transported electron. This shows that competing relaxation channels for the qubit, such as
phonon emission, are relevant in inelastic processes even in the presence of the resonator \cite{Fujisawa1998, Hohenester2009, Liu2014f, Gullans2015}. Irrespective of the details of the non-radiative relaxation mechanism we expect the rate of photon emission into the resonator to be limited by the ratio of the qubit dephasing rate to the cavity linewidth $\kappa/\gamma_\phi\sim0.01$.

We investigate emission for interdot tunnel coupling rates up to $2t/h=7.4$~GHz. Representative examples are shown in Fig.~\ref{fig:emission}(c). We find that the emission power at $-\delta_\text{0}$ increases with $t$, while the separation between the resonances, indicated by vertical dashed lines, decreases. To analyze the resonances in emission in detail we subtract the background signal proportional to the current. A sum of two Gaussian lineshapes is fitted to the resonances to analyze the maximum power and resonant detuning [Fig.~\ref{fig:emission}(c)]. We define the zero detuning to be centered between the resonances. For details of the PSD measurement and the extraction of the tunnel coupling, refer to the supplemental material \cite{SuppMat}.

The extracted resonant detuning $|\delta_\text{0}|/h$ approaches zero as $2t$ approaches $h\nu_0$ [Fig.~\ref{fig:theory}(a)]. For $2t\ge h\nu_0$, we find a single resonance [Fig.~\ref{fig:emission}(c)], which decreases in power and eventually vanishes when $t$ is further increased.
For $2t/h<4$~GHz the emission power at $-\delta_\text{0}$ is below the noise level of our detection system for the chosen integration time, so that no points can be acquired in this region. The experimental data are well described by Eq.~\eqref{deltares} and deviations lie within the error margins [Fig.~\ref{fig:theory}(a)] where all parameters are directly determined from the experiment.

\begin{figure}
	\centering
\includegraphics[width=8.6cm]{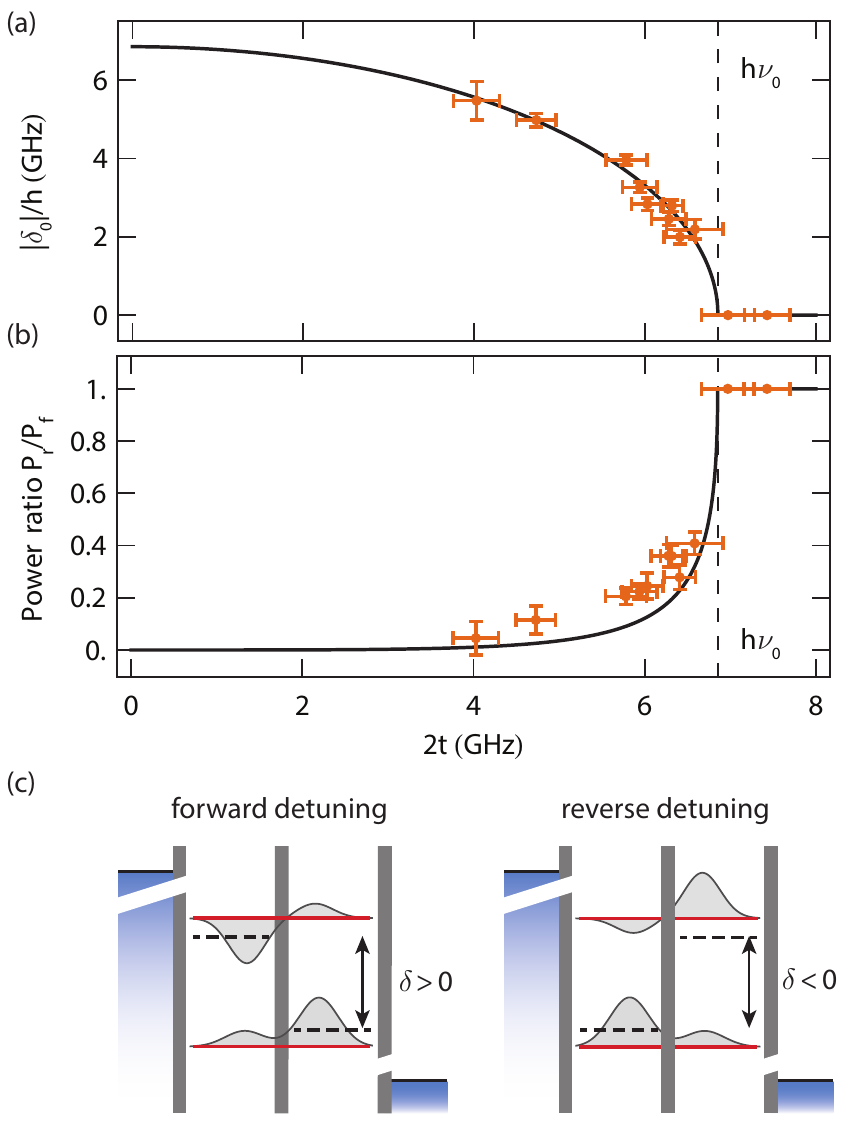}
	\caption{(a) Resonant detuning $|\delta_\text0|$ and (b) power ratio $P_\text{r}/P_\text{f}$ as a function of tunnel coupling $2t$. The data are compared to expressions \eqref{deltares} and \eqref{ratio} (solid curves). Error bars in $|\delta_\text{0}|$ and $P_\text{r}/P_\text{f}$ are determined by fitting error of the Gaussian lineshapes shown in Fig.~\ref{fig:emission}(c), error bars in $2t$ result from the Jaynes-Cummings simulations of transmission measurements.
(c) Level diagrams depicting the wavefunctions of the hybridized dot states for positive and negative detuning. The dashed black lines represent the dot energy levels in the left-right ($\ket L$,$\ket R$) basis, the solid red lines represent the hybridized states and the shaded gray areas the wavefunctions of the bonding and antibonding state.}
	\label{fig:theory}
\end{figure}

For the negative resonance condition $-\delta_\text{0}$, the detuning of the electronic states is reversed with respect to the source-drain bias [Fig.~\ref{fig:theory}(c)]. 
In this configuration, transport is expected to be blocked if the hybridization of the occupation states $\ket{L}, \ket{R}$ of the left and right dots is not considered. The hybridized states are given by
\begin{align}
\ket e&=\cos(\theta/2)\ket L +\sin(\theta/2)\ket R,\\
\ket g&=\sin(\theta/2)\ket L -\cos(\theta/2)\ket R,
\end{align}
where $\theta$ is given by $\tan\theta=2t/\delta$. Considering the antibonding wavefunction $\ket{e}$, 
we find the electron in the left dot with probability $\alpha=\cos^2(\theta/2)$, which is a measure of the hybridization of the qubit states [Fig.~\ref{fig:theory}(c)]. 
Even for reverse detuning, electrons can therefore tunnel into the excited state leading to photon emission at $-\delta_\text{0}$.

We find that emission power for forward detuning $P_\text{f}$ is roughly independent of $t$ while photon emission for reverse detuning $P_\text{r}$ increases with $t$ [Fig.~\ref{fig:emission}(c)]. The power ratio $P_\text r/P_\text f$ approaches unity as $2t$ approaches $h\nu_0$ and the peaks merge [Fig.~\ref{fig:theory}(b)].

We model the behavior of the emitted power with increasing $t$ using a master equation approach. We describe the system by three possible states: qubit in the excited state $\ket{e}$ (excess electron in antibonding state), qubit in ground state $\ket{g}$ (excess electron in bonding state) and qubit in state $\ket 0$, which we define as the state with no excess electrons in the dots. This describes the system around the lower triple point. For the upper triple point, the third qubit state corresponds to two extra electrons in the dots leading to the same result.
We arrive at the following rate equation for the occupation probabilities of the three qubit states \cite{Ou-Yang2010}: 
\begin{equation}
\left \{
\begin{array}{l r r r r r}
\dot p_0=&-\Gamma_\text L			p_0  	&+&\alpha\Gamma_\text R p_g &+&(1-\alpha) \Gamma_\text R p_e \\
\dot p_g=&(1-\alpha) \Gamma_\text L		p_0		&-&\alpha\Gamma_\text R p_g && \\
\dot p_e=&\alpha \Gamma_\text L	p_0 	 	&&								&-&(1-\alpha) \Gamma_\text R p_e.
\end{array}
\right.
\end{equation}
Here, we neglect the relaxation rate $\gamma_1$ from the excited to the ground state. This is justified because we estimate $\gamma_1\approx 100~\text{MHz}$ \cite{Fujisawa2002}, which is significantly smaller than the tunneling rates to the leads $\Gamma_L\approx\Gamma_R$. The same holds for the resonator-dot coupling $g\approx11$~MHz. We confirmed that including $\gamma_1$ in the rate equation does not significantly affect the power ratio even for $\gamma_1\gg 100$~MHz. 

Because the resonant emission signal is generated in transitions from the excited state
we expect the emission power to be proportional to $p_e(\delta)$, which is obtained using the steady-state condition $\dot p=0$. This yields the emission power ratio 
\begin{align}
\frac {P_{\text r}}{P_{\text f}}
=\frac{p_e(-\delta_{\text 0})} {p_e(+\delta_{\text 0})}
=\frac{(h\nu_0)^2-2t^2-h\nu_0\delta_\text{0}}{(h\nu_0)^2-2t^2+h\nu_0\delta_\text{0}}.
\label{ratio}
\end{align}

The experimental results are well approximated by this model [black solid line in Fig.~\ref{fig:theory}(b)]. There is, however, a systematic deviation, such that the power ratios in all experimental data are slightly higher than the model predicts. We can account for this deviation by considering energy dependent tunnel rates $\Gamma_\text L$ and $\Gamma_\text R$, see Ref.\ \onlinecite{SuppMat}. Note that, as for the resonant detuning [Fig.~\ref{fig:theory}(a)], there is no free fitting parameter.

In conclusion, we investigated photon emission from a biased double quantum dot dipole coupled to a microwave cavity. We clearly distinguish between background emission and resonant photon emission of the qubit generated in inelastic interdot transitions.
We found photon emission for both forward and reverse interdot detuning with respect to source-drain bias reflecting the distribution of wavefunctions in the hybridized electronic states. We modeled the results using a master equation approach for the double quantum dot and found good agreement with the experiment. In future experiments, microwave correlation function measurements could be used to investigate photon statistics \cite{Bozyigit2011,daSilva2010}. For the low photon rates we detect from single-electron interdot transitions in our experiments photon antibunching is predicted \cite{Xu2013a}.

\begin{acknowledgments}

We thank C.\ Eichler for valuable advice, A.\ \mbox{Hambitzer} for contributions to device fabrication, and C.\ Lang and Y. Salath\'e for their support. We gratefully acknowledge discussions with G. Sch\"on, A. Shnirman and P. Samuelsson. This work was financially supported by the Swiss National Science Foundation through the National Center of Competence in Research "Quantum Science and Technology" and by ETH Zurich.

\end{acknowledgments}

\newpage

\onecolumngrid

\begin{center}
\textbf{\large Supplemental material to\\``Microwave Emission from Hybridized States in a Semiconductor Charge Qubit"}\\
\vspace{\columnsep}
\text{A.~Stockklauser$^*$, V.~F.~Maisi, J.~Basset$^\dagger$, K.~Cujia, C.~Reichl, W.~Wegscheider, T.~Ihn, A.~Wallraff, K.~Ensslin}\\
\textit{Department of Physics, ETH Z\"urich, CH-8093 Zurich, Switzerland}\\
\text{(Dated: 24 July 2015)}
\end{center}

\vspace{\columnsep}

\twocolumngrid

\beginsupplement

\maketitle

\section{Measurement details}

The emission experiment is performed at finite source-drain bias where the triple points of a charge stability diagram expand into finite bias triangles [Fig.~\ref{fig:FBTs}]. We measure the emitted power at the lower triangle and vary the interdot detuning by tuning the left and right plunger gate voltages along the $\delta$-axis indicated by the dashed purple line [Fig.~3(b)]. 
We additionally study emitted power vs.\ tunnel coupling [Fig.~3(c)] by tuning the center gate [Fig.~1(b)] voltage.

\subsection{Extraction of tunnel coupling}

\begin{figure}
	\centering
\includegraphics[width=8.6 cm]{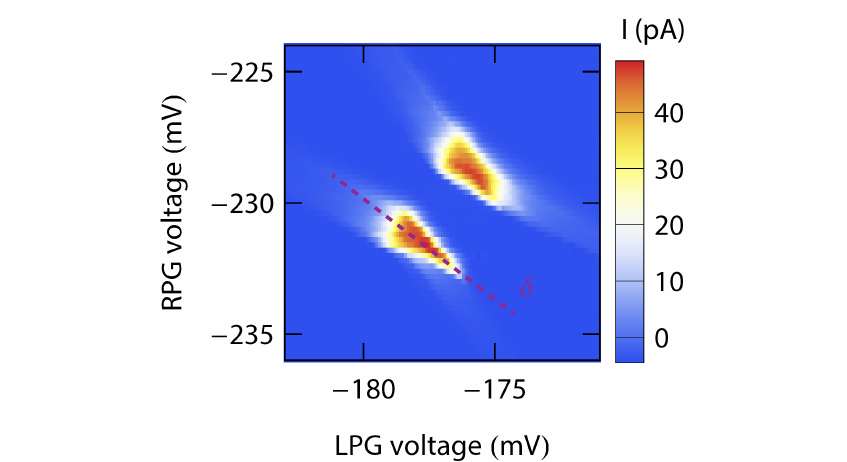}
	\caption{Finite bias triangles at $V_\text{SD}=-200~\mu$V showing the source-drain current $I$ vs. left and right plunger gate (LPG, RPG) voltage. The dashed purple line marks where the emission data is taken.}
	\label{fig:FBTs}
\end{figure}

\begin{figure}
	\centering
\includegraphics[width=8.6 cm]{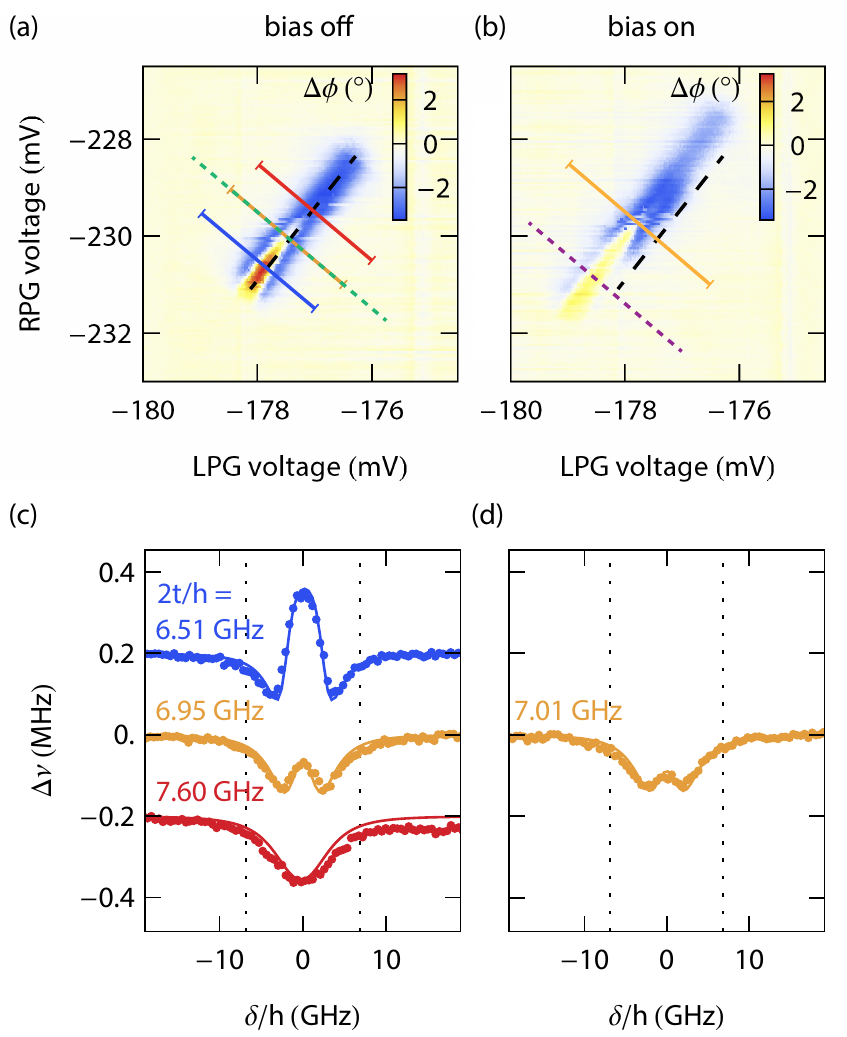}
	\caption{Measured phase shift $\Delta\phi$ at $V_\text{SD}=0$ (a) and $V_\text{SD}=-200~\mu$V (b). The dashed black line serves as a reference to illustrate the shift of the charge degeneracy line in the LPG/RPG plane when the bias is turned on. The dashed purple line indicates the gate voltage configuration where the emission data is acquired, while the dashed green line indicates where the bare tunnel coupling is extracted, i.e.\ where the curves shown in Fig.~\ref{fig:curves} are taken. (c), (d) Shift of the resonant frequency $\Delta\nu$ (colored dots) measured along the cuts indicated in (a) and (b). Solid lines are results of numerical simulations. In the left panel we observe that the tunnel rate $2t/h$ (insets) increases with plunger gate voltage. We compare the middle (orange) cut between bias off and on and see that $2t$ changes by $\sim60$~MHz in this case.}
	\label{fig:biasshift}
\end{figure}

For each center gate voltage, the tunnel coupling energy is extracted individually from microwave transmission measurements \cite{Frey2012}. 
Without source-drain bias, transmission measurements show a phase shift $\Delta\phi$ of the output microwave signal in the region of the charge degeneracy line connecting the triple points [Fig.~\ref{fig:biasshift}(a)]. At finite bias, the phase shift extends further 
along the base lines of the finite bias triangles.
We aim at extracting the interdot tunnel coupling $t$ of the biased system at the lower finite bias triangle corresponding to the conditions of the emission measurements [purple line in Fig.~\ref{fig:FBTs} and Fig.~\ref{fig:biasshift}(b)]. However, we observe that the charge transport occuring within the finite bias triangles changes the phase and frequency response of the resonator compared to the unbiased system. We therefore extract the tunnel coupling using the Jaynes-Cummings description from measurements without source-drain bias where there is no transport through the system. 

The tunnel coupling between the dots is not constant along the charge degeneracy line but decreases for more negative plunger gate voltages which is observed in the phase shift [Fig.~\ref{fig:biasshift}(a)]. This is due to the effect of the cross-coupling of the plunger gates on the central tunnel barrier and thereby on $t$. We can quantify the change of tunnel coupling by measuring full transmission spectra vs.\ detuning for multiple cuts through the charge degeneracy line [Fig.~\ref{fig:biasshift}(a) and (c)] and find that $t$ decreases linearly with plunger gate voltage. 

\begin{figure}
	\centering
\includegraphics[width=7.5 cm]{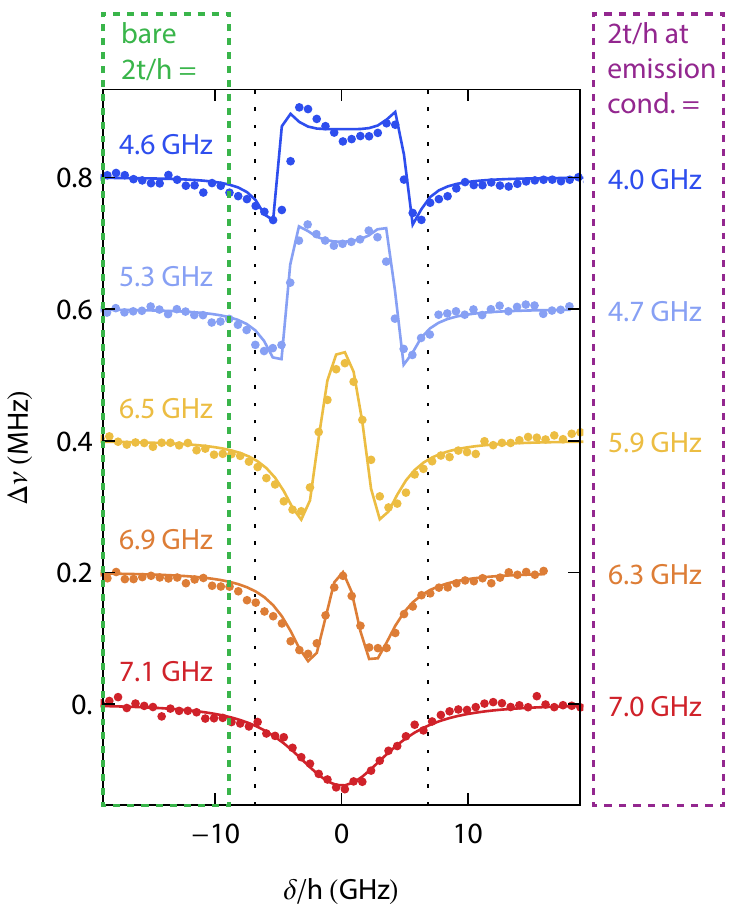}
	\caption{Measured frequency shift $\Delta\nu$ (colored dots) and numerical simulation (solid lines) to extract $t$.
The bare values of $2t/h$ extracted from the curves are indicated in the dashed green box on the left. The values given in the dashed purple box on the right take into account the shifts described in the text and are valid at the conditions of the emission experiments presented in the main text. The other parameters of the simulation are the coupling strength $g/2\pi=11$~MHz and decoherence $\gamma/2\pi=[500, 600, 800, 1000, 1000]$~MHz (from top to bottom).
}
	\label{fig:curves}
\end{figure}

In the unbiased case the cavity frequency shift disappears at the triple points, so that the Jaynes-Cummings Hamiltonian cannot model the frequency shift $\Delta\nu$ and $t$ has to be extracted between the triple points [dashed green line, Fig.~\ref{fig:biasshift}(a)].  
For the emission measurements presented in Fig.~3(c), the bare interdot tunnel coupling is extracted from the frequency shifts $\Delta\nu$ shown in Fig.~\ref{fig:curves} and the values are given in the dashed green box.
We extrapolate the linear relation between $t$ and the plunger gate voltages to obtain $t$ at the triple point.

Additionally, we need to take the effect of the bias on the tunnel coupling into account. Applying a bias of $V_\text{SD}=-200~\mu$V shifts the charge degeneracy line slightly in the LPG/RPG plane leading to a change of $\approx 50$~MHz in $2t/h$ for points on the charge degeneracy line [cf.\ orange curves in Fig.~\ref{fig:biasshift}(c) and (d)]. 
With this correction we find the shifted $2t/h$ for the parameter settings of the emission experiment, which are listed in the dashed purple box in Fig.~\ref{fig:curves}.

The simulation based on the Jaynes-Cummings model yields $2t/h$ up to a systematic uncertainty between 50 and 200 MHz, depending on the tunnel coupling. An additional error is introduced by the linear extrapolation described above. Finally, the correction for the bias shift also entails a small additional uncertainty. The error bars plotted in Fig.~4(a) and (b) include these three error sources.

\subsection{Power spectral density}

The field quadratures I and Q of the emitted microwave radiation are measured in a heterodyne detection scheme. Fast signal processing with a field programmable gate array (FPGA) 
gives access to correlations of the emitted photons. 
We compute the power spectral density by taking the Fourier transform of the first-order auto-correlation function of the complex signal amplitude $S=I+iQ=Ae^{i\phi}$ \cite{Lang2011}.

We need to resolve a small photon signal on the order of $10^{-2}~\text{Hz}^{-1}s^{-1}$ generated by the biased double quantum dot. 
To distinguish this signal from the noise, we subtract zero-bias reference data. 
For this purpose, we alternate between measurements with the bias turned on and off in this case taking 40 repetitions.
After averaging the measurement (bias on) and the reference (bias off) data, we obtain the PSDs $S_\text{on}$ and $S_\text{off}$. Subtracting the two yields the power spectral density of the signal modes in which we are interested: $\text{PSD}=S_\Delta=S_\text{on}-S_\text{off}.$
Details about the measurement technique and signal processing can be found in Ref.~\onlinecite{Lang2014}.

The PSD plotted in Fig.~3(a) is scaled to represent the number of photons emitted from the resonator per Hz and per second. To obtain this scaling
, the number of noise photons at the input of the parametric amplifier was determined using the AC Stark shift observed in a superconducting qubit measurement with a similar detection chain in the same cryostat.

\section{Rate equation calculations}

In the model presented in the main text we neglect the decay rate $\gamma_1$ and assume energy independent tunnel barriers. This yields the rate equation $\dot p=Mp$ for the probabilities to find the system in one of its three states, where $p=(p_0, p_g, p_e)^T$ and 
\[ M=\left( \begin{array}{ccc}
-\Gamma_{\rm L} & \alpha\Gamma_{\rm R} & (1-\alpha) \Gamma_{\rm R} \\
(1-\alpha) \Gamma_{\rm L} & -\alpha\Gamma_{\rm R} & 0 \\
\alpha \Gamma_{\rm L}  & 0 &-(1-\alpha) \Gamma_{\rm R}\end{array} \right).\]

We calculate $p_\text{e}(\delta)$ using the steady-state condition $\dot p=0$ and $p_0+p_\text g+p_\text e=1$ and find
$$p_\text{e}(\delta)=\frac{\Gamma_\text{L} \left(2t^2+\delta \left( \delta+
\sqrt{4t^2+\delta^2}\right)\right)}{2t^2(\Gamma_\text R+2\Gamma_\text L)+2\Gamma_\text L\delta^2},$$

\begin{align}
p_\text{e}(\pm\delta_\text{0})=\frac{\Gamma_\text{L} \left((h\nu_0)^2-2t^2\pm 
h\nu_0\delta_\text0\right)}{2t^2(\Gamma_\text R-2\Gamma_\text L)+2\Gamma_\text L(h\nu_0)^2}.
\label{pe}
\end{align}
This expression describes the $t$-dependence of the power emitted at each resonance condition $\pm\delta_0$ and is indicated with dashed lines in Fig.~\ref{fig:heightscorr}(a). We assume $\Gamma_\text L=\Gamma_\text R$ and scale the curves to agree with the measured emission power. 
Experimentally, we observe that the emitted power at $-\delta_\text{0}$ (blue) increases with $t$, while the emission power at $+\delta_\text{0}$ (red) hardly depends on $t$, which clearly deviates from the behavior predicted by Eq.~\eqref{pe}.
A possible $\delta$-dependence of the decay rate from excited to ground state related to 
photon emission into the resonator is not taken into account.
Note that the peaks are merged for the last two data points, which are therefore omitted in the figure.

This deviation is also apparent when we consider the power ratio 
$$\frac {P_\text r}{P_\text f}
=\frac{p_\text{e}(-\delta_\text{0})} {p_\text{e}(+\delta_\text{0})}
=\frac{(h\nu_0)^2-2t^2-
h\nu_0\delta_\text{0}}{(h\nu_0)^2-2t^2+ 
h\nu_0\delta_\text{0}},$$
which is plotted in black in Fig.~\ref{fig:heightscorr}(b). No assumptions for the tunnel rates to the leads ($\Gamma_\text L$, $\Gamma_\text R$) and no scaling factor enter this expression.

\subsection{Energy dependent tunnel rates to the leads}

We can generalize the model by allowing an energy dependence of the tunnel rates $\Gamma_\text L$ and $\Gamma_\text R$, i.e.\ the tunnel rates into and out of the excited state $\Gamma_\text L^e$, $\Gamma_\text R^e$ differ from those into and out of the ground state $\Gamma_\text L^g$, $\Gamma_\text R^g$ \cite{MacLean2007}. The energy dependence can arise from changes in the tunnel coupling or the density of states in the leads. We find the generalized transition matrix: 
\[ \tilde M=\left( 
\begin{array}{ccc}
-(1-\alpha)\Gamma_\text L^g-\alpha \Gamma_\text L^e & \alpha\Gamma_\text R^g & (1-\alpha) \Gamma_\text R^e \\
(1-\alpha) \Gamma_\text L^g & -\alpha\Gamma_\text R^g & 0\\
\alpha \Gamma_\text L^e  & 0 &-(1-\alpha) \Gamma_\text R^e
\end{array} \right)\] 
The steady state condition $\dot p=\tilde M p=0$ now yields a modified excited state occupation probability $\tilde p_\text{e}(\delta,\Gamma_\text L^g,\Gamma_\text L^e,\Gamma_\text R^g,\Gamma_\text R^e)$.
A first assumption could be that the left and right barriers are both energy dependent in the same way, i.e.\ $\Gamma_\text L^e=c\Gamma_\text L^g$ and $\Gamma_\text R^e=c\Gamma_\text R^g$ with some factor c typically $>1$. This, however, does not influence the theoretically predicted power ratio.

\begin{figure}
	\centering
\includegraphics[width=8.6 cm]{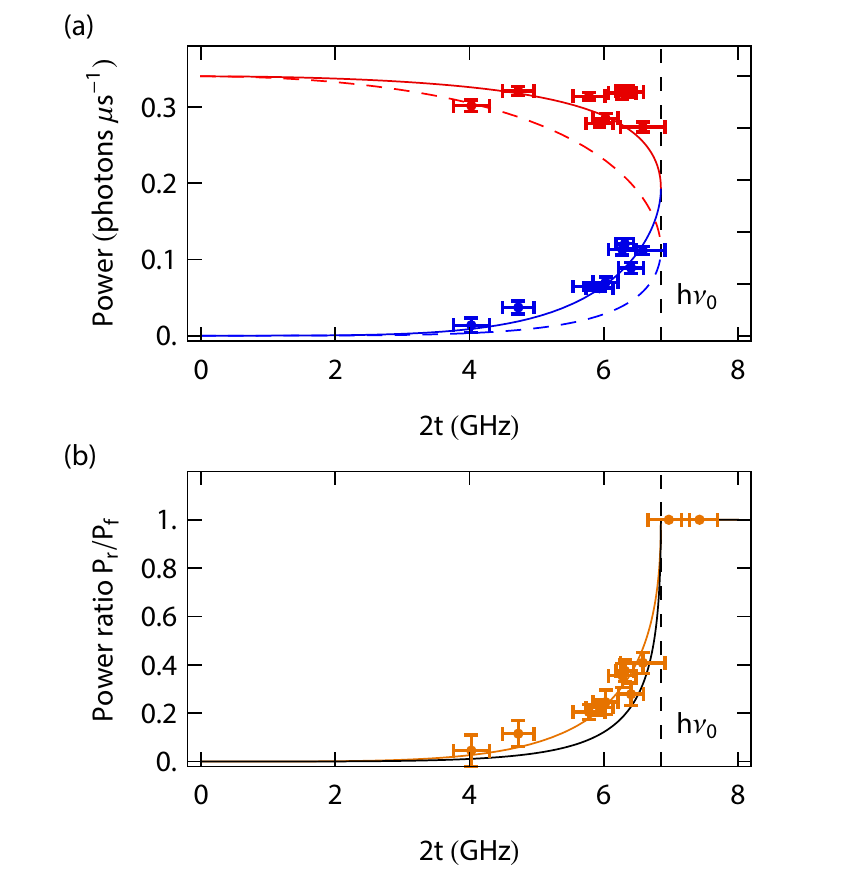}
	\caption{(a) Emitted power at $-\delta_\text{0}$ (blue) and $+\delta_\text{0}$ (red). The dashed lines are the theoretical expressions given by Eq.~\eqref{pe}, where we assumed $\Gamma_\text L=\Gamma_\text R$ and a scaling factor. The solid lines are plots of $\tilde p_\text{e}(-\delta_\text{0})$ (blue) and $\tilde p_\text{e}(+\delta_\text{0})$ (red) assuming  $\Gamma=\Gamma_\text L^g=\Gamma_\text R^e=\Gamma_\text R^g$ and  $\Gamma_\text L^e=c\Gamma$ with $c=2.6$. (b) Power ratio $P_\text r/P_\text f$. The solid black line shows the theoretical ratio obtained with energy independent tunnel barriers [cf.\ Fig~4(b)], while the solid orange line was obtained from the generalized model assuming energy dependent tunnel barriers. }
	\label{fig:heightscorr}
\end{figure}

We can correct for the deviation in Fig.~\ref{fig:heightscorr}(b) if we assume that only the tunnel rate from the left lead increases with energy: $\Gamma_\text L^e>\Gamma_\text L^g$. More specifically, we assume $\Gamma=\Gamma_\text L^g=\Gamma_\text R^e=\Gamma_\text R^g$ and  $\Gamma_\text L^e=c\Gamma$. We fit the data to
the modified power ratio yielding $c=2.6$. The order of magnitude of this energy dependence seems realistic considering the findings of other transport experiments \cite{Amasha2008, Roessler2015}. We assume that the origin for the energy dependence lies in a modulated density of states due to gate geometry and disorder \cite{Roessler2015}. 
In the measured device the left and right side gate influence the double dot confinement potential very differently, which is not expected from the device geometry.
This could be the reason for the left/right asymmetry of the tunnel rates to the leads.

The resulting modified expressions $\tilde p_\text{e}(\pm\delta_0)$ for the emitted power and the power ratio $P_\text r/P_\text f$ are shown in Fig.~\ref{fig:heightscorr} and the agreement with the experimental data is considerably improved. Simply assuming $\Gamma_\text L>\Gamma_\text R$ does not influence the power ratio, the asymmetric energy dependence is therefore necessary to account for the deviation.

\subsection{Resonance linewidth}

\begin{figure}
	\centering
\includegraphics[width=8.6 cm]{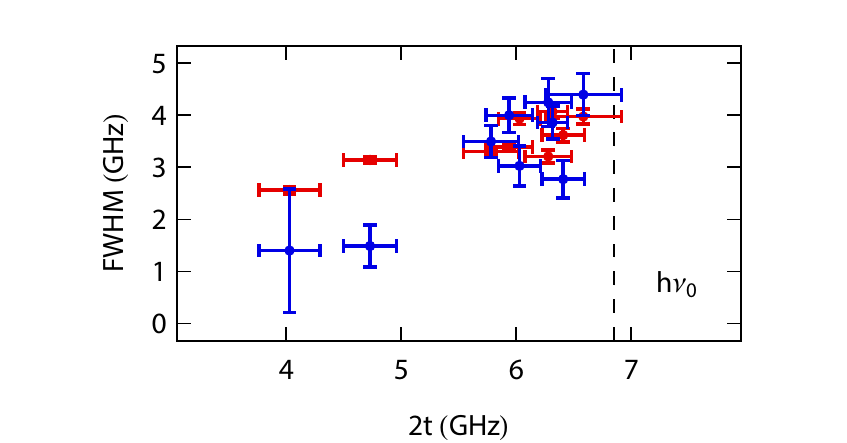}
	\caption{Full width at half maximum (FWHM) of the emission resonances at $-\delta_\text{res}$ (blue) and $+\delta_\text{res}$ (red).}
	\label{fig:FWHM}
\end{figure}

Further information about the emission process can be gained from analyzing the linewidth of the emission resonances. The Gaussian peaks fitted to the data in Fig.~3.(c) have a full width at half maximum (FWHM) between 1 and 5~GHz increasing with $t$ [Fig.~\ref{fig:FWHM}].

This linewidth can be set in relation to the energy broadening of the electronic dot states. We assume that the resonances of the electronic states are Lorentzian with a FWHM $\Gamma/2$. 
The emission signal is described by a convolution of the two states involved in the transition, which yields a Lorentzian in $\nu_q$ with linewidth $\Gamma$:
$$P(\nu_q)=A(\delta)\frac{\Gamma/2}{(\nu_q(\delta)-\nu_0)^2+(\Gamma/2)^2}.$$
The emission amplitude $A(\delta)$ depends on the detuning between the dot states and is in our model proportional to the excited state occupation probability: $A(\delta)\propto p_\text e(\delta)$.
The resonator linewidth $\kappa/2\pi = 3.3$~MHz is much smaller than $\Gamma/2$, so that the resonator response can be approximated as a $\delta$-function and neglected in this consideration.

We are interested in the emission resonances at constant $t$ as a function of detuning $\delta$:
\begin{align}
P(\delta/h)=A(\delta)\frac{\Gamma/2}{(\sqrt{(2t)^2+\delta^2}/h-\nu_0)^2+(\Gamma/2)^2}.
\label{lorentz}
\end{align}
This is not Lorentzian in $\delta$ but yields two peaks symmetric in $\delta$, explaining the two resonances we observe in emission. 

By fitting Eq.~\eqref{lorentz} to the measured emission resonances, we find $\Gamma=1.5\pm0.4$~GHz. This corresponds to a width $\Gamma/2=0.75\pm0.2$~GHz of the electronic quantum dot states. 
We assume that the main contributions to the broadening of the electronic states are the double dot decoherence and the coupling to the leads. In the emission measurements the decoherence lies between 0.5 and 1~GHz while the coupling rate to the leads is $\Gamma_L\approx\Gamma_R\approx1$~GHz. Because both of these contributions are of the same order of magnitude we cannot distinguish their individual effects. Investigating the effect of decoherence on the emission linewidth would, however, be an interesting subject of further study.

\end{document}